\documentclass[conference]{IEEEtran}
\usepackage[T1]{fontenc}
\usepackage{txfonts}
\usepackage{graphicx}
\newcommand{\ket}[1]{|#1\rangle}
\newcommand{\ketbra}[1]{|#1\rangle\langle #1 |}
\newcommand{\bra}[1]{\langle #1 |}

\newcommand{\Tr}{\mathrm{Tr}}

\begin{document}
\title{Key Rate of the B92 Quantum Key Distribution Protocol
with Finite Qubits\thanks{%
Accepted for presentation in 2015 IEEE International
Symposium on Information Theory, Hong Kong,
June 14--19, 2015.}}
\author{Hiroaki Sasaki\IEEEauthorrefmark{1},
Ryutaroh Matsumoto\IEEEauthorrefmark{1}, and
Tomohiko Uyematsu\IEEEauthorrefmark{1}\\
{\IEEEauthorrefmark{1}Department of Communications and Computer Engineering,
Tokyo Institute of Technology,
Japan.}}

\maketitle

\begin{abstract}
The key rate of the B92 quantum key distribution protocol
had not been reported before this
research when the number of qubits is finite.
We compute it by using the security analysis framework proposed
by Scarani and Renner in 2008.
\end{abstract}


\section{Introduction}
The B92 quantum key distribution (QKD) protocol
\cite{bennett92qkd} has remained less popular than the
famous BB84 protocol \cite{bennett84},
while both protocols provide the unconditional security.
One plausible reason for the unpopularity is that
the B92 is weaker to the channel noise than the BB84.
Specifically, the BB84 with the standard one-way
information reconciliation can generate secure key
over the depolarizing channel at  depolarizing rate
16.5\%, while
the previous  security analyses of the B92
cannot guarantee the secure key generation at
depolarizing rate 3.5\% \cite{tamaki03},
3.7\% \cite{genericqkd} or 4.2\% \cite{renner05}.
By using the security analysis framework introduced by
Renner in 2005 \cite{rennerphd},
we improved the maximal tolerable depolarizing rate
to 6.5\% \cite{matsumoto13b92}.

All of the above analyses \cite{genericqkd,matsumoto13b92,renner05,tamaki03}
assumed the infinite number of qubits in the protocol,
and derived the asymptotic key rates.
On the other hand, in practice the number of qubits used in a protocol
is always finite.
However, before this research,
the key rates with finite qubits in the B92 protocol
had not been reported,
as far as the authors know.
In this paper, we report the key rates with finite qubits,
based on the analytic framework
introduced by Scarani and Renner \cite{scarani08prl}
and our previous researches \cite{matsumoto13b92,sano10}.
We stress that the assumption in our paper is the same as
\cite{scarani08prl}, and in particular we assume the
collective attack instead of the coherent attack.

\section{New Security Analysis of the B92 Protocol with
Finite Qubits}
In this section, we present
a new formula for the key rate of the B92 protocol
with finite qubits,
based on previous researches \cite{matsumoto13b92,sano10,scarani08prl}.
The following description has some overlap with our previous research
improving the asymptotic key rate of the B92 \cite{matsumoto13b92}.
Firstly, we fix notations.
Let $\{\ket{0}$, $\ket{1}\}$ be some fixed orthonormal
basis of a qubit.
In the B92 protocol \cite{bennett92qkd},
Alice sends the quantum state
\begin{equation}
\ket{\varphi_j} = \beta\ket{0}+(-1)^j\alpha\ket{1}, \label{eq:alpha}
\end{equation}
for $j=0,1$, where $\beta = \sqrt{1-\alpha^2}$,
and $0 < \alpha< 1/\sqrt{2}$.
For convenience of presentation, we also define
\[
\ket{\bar{\varphi}_j} =\alpha\ket{0} - (-1)^j \beta\ket{1}.
\]
We can see that $\{\ket{\varphi_j}$, $\ket{\bar{\varphi}_j}\}$ forms
an orthonormal basis of a qubit.

On the other hand,
we can express a qubit channel as follows.
Define the three Pauli matrices $\sigma_x$, $\sigma_y$,
and $\sigma_z$ as usual.
Then a qubit density matrix $\rho$ can be expressed
as \cite{chuangnielsen}
\[
\rho = \frac{1}{2}\left(
I + x \sigma_x + y \sigma_y + z \sigma_z\right),
\]
where $x,y,z\in \mathbf{R}$ and
$x^2 + y^2 + z^2 \leq 1$.
The vector $(x,y,z)$ is called a Bloch vector.
The qubit channel $\mathcal{E}_B$ from Alice
to Bob can be expressed \cite{fujiwara99} as a map between
Bloch vectors by
\begin{equation}
\left(\begin{array}{c}
z\\
x\\
y
\end{array}\right)
\mapsto
R \left(\begin{array}{c}
z\\
x\\
y
\end{array}\right) + \vec{t},\label{eq:channelparameter}
\end{equation}
where
\begin{equation}
R = \left(
\begin{array}{ccc}
R_{zz}&R_{zx}&R_{zy}\\
R_{xz}&R_{xx}&R_{xy}\\
R_{yz}&R_{yx}&R_{yy}
\end{array}
\right), \quad
\vec{t} = 
\left(\begin{array}{c}
t_z\\
t_x\\
t_y
\end{array}\right). \label{eq:parameters}
\end{equation}

Define
\[
\ket{\Psi}
= \frac{\ket{0}_A\ket{\varphi_0}_B + \ket{1}_A\ket{\varphi_1}_B}{\sqrt{2}}.
\]
As in \cite{tamaki03},
we also define the four POVM 
\begin{eqnarray}
F_0 &=& \ket{\bar{\varphi}_1}\bra{\bar{\varphi}_1}/2,\label{povm1}\\
F_1 &=& \ket{\bar{\varphi}_0}\bra{\bar{\varphi}_0}/2,\\
F_{\bar{0}} &=& \ket{{\varphi}_1}\bra{{\varphi}_1}/2,\\
F_{\bar{1}} &=& \ket{{\varphi}_0}\bra{{\varphi}_0}/2.\label{povm4}
\end{eqnarray}
In \cite{tamaki03}, the measurement outcomes corresponding to
$F_{\bar{0}}$ and $F_{\bar{1}}$ was not distinguished.
We distinguish them for better channel estimation.

After passing the quantum channel $\mathcal{E}_B$
from Alice to Bob,
$\ket{\Psi}\bra{\Psi}$ becomes
\begin{equation}
\rho_{1,AB} = (I \otimes \mathcal{E}_B) \ket{\Psi}\bra{\Psi}.\label{eq:afterchannel}
\end{equation}
In a quantum key distribution protocol,
the state change $\mathcal{E}_B$ is caused by Eve's cloning of the
transmitted qubits to her quantum memory. The content of Eve's quantum
memory is mathematically described by the purification
$\ket{\Phi_{1,ABE}}$ of $\rho_{1,AB}$.
Let $\rho_{1,ABE} = \ket{\Phi_{1,ABE}}\bra{\Phi_{1,ABE}}$.

In addition to Eve's quantum memory,
she also knows the content of public communication over the
classical public channel between Alice and Bob.
For each transmitted qubit from Alice to Bob,
the public communication consists of $1$-bit information
indicating whether Bob excludes his received qubit for generating
the final secret key or not.
We also have to take it into account. We shall represent the
public communication by a classical random variable $P$
that becomes $1$ if Bob excludes his qubit and $0$ otherwise.
So, $P=0$ when Bob's measurement outcome is $F_0$ or $F_1$,
and $P=1$ when Bob's measurement outcome is $F_{\bar{0}}$ or $F_{\bar{1}}$.

On the other hand, in the B92 protocol,
Bob performs the measurement specified by Eqs.\ (\ref{povm1})--(\ref{povm4}).
Alice and Bob use their qubit for generation of the
final secret key only if its measurement outcome
is $F_0$ or $F_1$. Otherwise it is excluded from the
key generation.
This is mathematically equivalent to set Alice's bit to $0$ if
the measurement outcomes is $F_{\bar{0}}$ or $F_{\bar{1}}$.
Therefore, from Eve's perspective on Alice's classical bit,
the joint state between Alice and Bob after the selection by measurement
outcomes is equivalent to
\begin{eqnarray*}
\rho_{2,ABEP} &=& (I_{A}\otimes \sqrt{F_0} \otimes I_E\rho_{1,ABE} I_{A}\otimes \sqrt{F_0}\otimes I_E\\ && + I_{A}\otimes \sqrt{F_1}\otimes I_E\rho_{1,ABE} I_A\otimes \sqrt{F_1}\otimes I_E)\otimes \ket{0}_P\bra{0}_P \\
&& +
\ket{0}_A\bra{0}_A\otimes (\sqrt{F_{\bar{0}}}\otimes I_E\Tr_A[\rho_{1,ABE}]\sqrt{F_{\bar{0}}}\otimes I_E\\ &&
 + \sqrt{F_{\bar{1}}}\otimes I_E\Tr_A[\rho_{1,ABE}]\sqrt{F_{\bar{1}}}\otimes I_E)\otimes \ket{1}_P\bra{1}_P. \label{eq:afterselection}
\end{eqnarray*}
Observe that the state change from $\rho_{1,ABE}$ to
$\rho_{2,ABEP}$ is a trace-preserving completely positive map.

In order to calculate the key rate, we need to consider
Eve's ambiguity on Alice's classical bit \cite{renner05,rennerphd}
defined as follows. 
Let
\[
\rho_{2,XEP} = \sum_{j=0,1} \ket{j}_A\bra{j}_A\otimes I_{EP} \mathrm{Tr}_B[\rho_{2,ABEP}] \ket{j}_A\bra{j}_A\otimes I_{EP}.
\]
Eve's ambiguity on Alice's classical bit $S(X|EP)$ is defined as
\begin{equation}
S(X|EP) = S(\rho_{2,XEP}) - S(\rho_{2,EP}),\label{eq:sxe}
\end{equation}
where $\rho_{2,EP} = \Tr_A[\rho_{2,XEP}]$, and $S(\cdot)$ denotes
the von Neumann entropy.

In order to calculate the amount of public communication required for
information reconciliation, we define the joint random variables
$(X',Y')$ as
\begin{eqnarray}
X'&=& j\textrm{ if the transmitted qubit is }\ket{\varphi_j}, \nonumber\\
Y'&=& k \textrm{ if the measurement outcome is }F_k,\label{eq:y}
\end{eqnarray}
under the condition that the measurement outcome is
either $F_0$ or $F_1$.
Observe the difference between $X$ and $X'$.
$X'$ is not defined  but $X$ is defined to be $0$
when Bob's measurement outcome is either 
$F_{\bar{0}}$ or $F_{\bar{1}}$.

We shall show the key rate per single transmitted
qubit that is neither announced for the channel estimation nor
excluded due to the measurement outcome being $F_{\bar{0}}$ or $F_{\bar{1}}$.
Note that Eq.\ (\ref{eq:sxe}) is Eve's ambiguity per a qubit
that is not announced for the channel estimation but \emph{can be
discarded}.
The probability of the measurement outcome being
$F_0$ or $F_1$ is
\[
\Tr[\rho_{1,AB} (I_A \otimes (F_0 + F_1))].
\]
So we can see that Eve's ambiguity  per single transmitted
qubit that is neither announced for the channel estimation nor
discarded is
\[
\frac{S(X|EP)}{\Tr[\rho_{1,AB} (I \otimes (F_0 + F_1))]}.
\]
By \cite{renner05,rennerphd} the \emph{asymptotic} key rate is
\begin{equation}
\frac{S(X|EP)}{\Tr[\rho_{1,AB} (I \otimes (F_0 + F_1))]} - H(X'|Y').\label{eq:keyrate}
\end{equation}
The above analysis is almost the same as our previous one
\cite{matsumoto13b92} for the asymptotic key rate
assuming the infinite number of qubits.

Note that the above formula (\ref{eq:keyrate}) assumes that
Alice and Bob know the channel between them.
In the BB92 protocol,
we cannot estimate all the parameters of the channel,
even if we assume infinitely many qubits in the protocol.
We can only estimate part of them.
In addition to that,
because the number of qubits in the protocol is finite,
there must be statistical errors.

To handle the finiteness of qubits,
Scarani and Renner \cite{scarani08prl}
used the interval estimation of channel parameters
($R$ and $\vec{t}$ of (\ref{eq:parameters}) in our study).
In contrast to the more popular point estimation,
by using
statistical samples,
interval estimation gives a set of parameters
that contains true parameters with high probability $1-\epsilon_{\mathrm{PE}}$.
By using the results in \cite{scarani08prl},
the key rate of the B92 protocol can be computed as
\begin{equation}
  r = \min_{(R,\vec{t}) \in \Gamma(\epsilon_{\mathrm{PE}})} S(X|EP) -  H(X'|Y') - \Delta  /n , \label{eq:finiterate}
\end{equation}
where $\Gamma(\epsilon_{\mathrm{PE}})$ is a confidence region given by
an interval estimation procedure with the confidence level $\geq 1 - \epsilon_{\mathrm{PE}}$, $\Delta$ is as defined in \cite[Eq.\ (5)]{scarani08prl}, and
$n$ is the number of the qubits to which Alice and Bob apply the privacy
amplification.

To compute the rate (\ref{eq:finiterate}), there are two remaining
tasks, namely (a) computation of $\Gamma(\epsilon_{\mathrm{PE}})$,
and (b) computation of $\min_{(R,\vec{t}) \in \Gamma(\epsilon_{\mathrm{PE}})} S(X|EP)$.
Task (b) is performed by using the convex optimization method
\cite{boyd04} as done in our previous researches \cite{matsumoto13b92,sano10}.
For convex optimization, the confidence region $\Gamma(\epsilon_{\mathrm{PE}})$
must be a convex set that can be easily handled by a mathematical
software, like Mathematica.
In \cite{sano10}, such a convex confidence region was introduced
for the BB84 protocol by using the KL divergence.
We shall define $\Gamma(\epsilon_{\mathrm{PE}})$ also by using the
KL divergence.

In the conventional researches \cite{genericqkd,matsumoto13b92,renner05,tamaki03},
their channel estimation procedures classified
Bob's measurement outcomes into three categories,
namely, $F_0$, $F_1$ and the inconclusive ($F_{\bar{0}}$ or $F_{\bar{1}}$).
In this research, we propose to distinguish 
$F_{\bar{0}}$ and $F_{\bar{1}}$ for better estimation accuracy.
On the other hand, the conventional estimation procedures 
did not distinguish which $\ket{\varphi_0}$ or $\ket{\varphi_1}$
produced Bob's measurement outcome.
We also propose to distinguish Alice's transmitted qubits
$\ket{\varphi_0}$ and $\ket{\varphi_1}$
in channel estimation.

By the above consideration,
the proposed channel estimation procedure has at least 8 kinds
of outcomes.
On the other hand,
the treatment of Bob's outcome $F_0$, $F_1$, 
$F_{\bar{0}}$ and  $F_{\bar{1}}$ is asymmetric,
because all of $F_{\bar{0}}$ and  $F_{\bar{1}}$
are disclosed to Alice and are used for channel estimation,
while parts of $F_0$ and  $F_1$ are kept 
secret for the secret key generation.
Because of 
this asymmetry,
the sum of 8 POVM operators corresponding the above 8 outcomes
does not become the $4\times 4$ identity matrix $I_{4 \times 4}$.
To make the sum equal to $I_{4 \times 4}$,
we include the outcome meaning the qubit kept secret
for secret key generation.
By $r_{pub}(0 < r_{pub} < 1)$ we denote the conditional
probability for a qubit being disclosed for channel estimation,
and the qubit is kept secret for secret key generation
with a probability $r_{pub}$.
We define the following 8 POVM operators:
\begin{eqnarray}
  E_0 &=& r_{pub} \ketbra{0_A} \otimes F_0 \\
  E_1 &=& r_{pub} \ketbra{0_A} \otimes F_1 \\
  E_2 &=&  \ketbra{0_A} \otimes F_{\bar{0}} \\
  E_3 &=&  \ketbra{0_A} \otimes F_{\bar{1}} \\
  E_4 &=& r_{pub} \ketbra{1_A} \otimes F_0 \\
  E_5 &=& r_{pub} \ketbra{1_A} \otimes F_1 \\
  E_6 &=&  \ketbra{1_A} \otimes F_{\bar{0}} \\
  E_7 &=&  \ketbra{1_A} \otimes F_{\bar{1}} \\
  E_8 &=& (1 - r_{pub}) I_{2\times 2} \otimes (F_0 + F_1).
\end{eqnarray}
The last operator $E_8$ corresponds to
the imaginary measurement outcome expressing
the non-disclosure of a qubit.

By this preparation of notations,
we can describe the proposed confidence region of the channel parameters.
Let $D(P\| Q)$ denotes the Kullback-Leibler divergence,
$\lambda(\rho_{1,AB})$ the theoretical probability distribution
of the 9 outcomes defined as
\[
\lambda_\infty(\rho_{1,AB}) = (\mathrm{Tr}[\rho_{1,AB}E_0], \ldots, \mathrm{Tr}[\rho_{1,AB}E_8]),
\]
and $\lambda_m$ the empirical distribution (i.e.\ relative frequencies)
of the 9 outcomes, where $m$ is the total number of qubits
transmitted including both disclosed and non-disclosed qubits.
Observe that
Alice and Bob can compute $\lambda_m$ in the protocol execution,
and their task is to estimate the channel parameters $(R, \vec{t})$.
The set
\begin{equation}
\{ (R,\vec{t}) \mid D(\lambda_m \| \lambda_\infty(\rho_{1, AB})) \leq \epsilon_{\mathrm{PE}},
(R,\vec{t}) \textrm{ defines a CP map}\} \label{eq:region}
\end{equation}
is a confidence region of $(R,\vec{t})$ 
with confidence level at least $1-\epsilon_{\mathrm{PE}}$,
by the well-known fact \cite[Theorem 11.2.1]{cover06}.
It is also well-known that the set of $(R,\vec{t})$
yielding a CP map is convex \cite{fujiwara99},
and $D(\cdot \| \cdot)$ is a convex function.
Therefore the set (\ref{eq:region}) is a convex set.
The above idea is similar to our previous research \cite{sano10} on BB84.
We have verified that the set (\ref{eq:region}) can be used
as $\Gamma(\epsilon_{\mathrm{PE}})$ in (\ref{eq:finiterate}).

The minimization in (\ref{eq:finiterate})
is just a convex optimization and can be done as follows.
Observe first that $S(X|EP)$ is a function of the channel 
parameters (\ref{eq:parameters}) of $\mathcal{E}_B$.
By the almost same argument as \cite[Remark 11]{watanabe08}
one sees that $S(X|EP)$ is a convex function of the channel 
parameters (\ref{eq:parameters}).
Moreover,
we see that the minimum of $S(X|EP)$ is attained 
when $R_{xy} = R_{yx} = R_{yz} = R_{zy} = t_y = 0$
by the almost same argument as \cite[Proposition 1]{watanabe08}.
Therefore, one can compute the minimization of $S(X|EP)$ by
the convex optimization \cite{boyd04}.



\begin{figure}[t!]
\includegraphics[angle=-90,width=\linewidth]{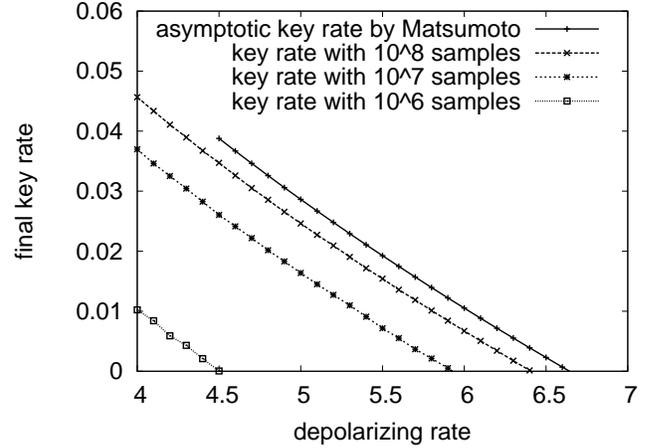}
\caption{Key rates with various depolarizing rates of the quantum channel}
\label{fig1}
\end{figure}

\begin{figure}[t!]
\includegraphics[angle=-90,width=\linewidth]{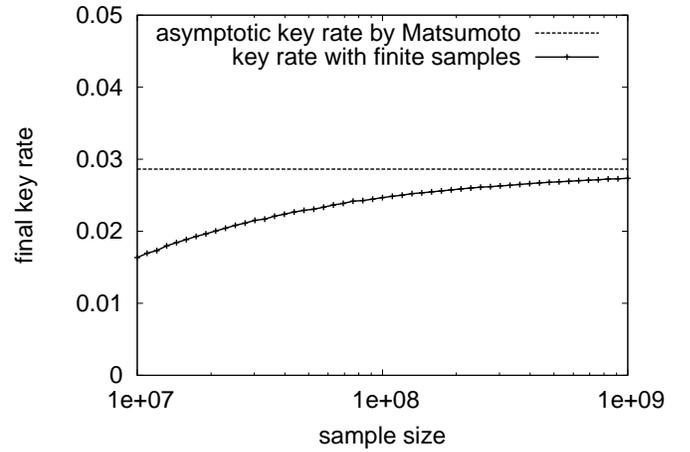}
\caption{Key rates with various sample sizes (depolarizing rate is 5\%)}
\label{fig2}
\end{figure}

\section{Numerical Result}
We consider the depolarizing channel $\mathcal{E}_q$
with
depolarizing rate $q$.
The definition of $q$ follows \cite{tamaki03}.
For a qubit density matrix $\rho$,
we have $\mathcal{E}_q(\rho) = (1-q) \rho + (q/2) I_{2 \times 2}$.
With such a channel $\mathcal{E}_q$,
$R$ and $\vec{t}$ in Eq.\ (\ref{eq:channelparameter})
are given by
\[
R = \left(
\begin{array}{ccc}
1-4q/3&0&0\\
0&1-4q/3&\\
0&0&1-4q/3
\end{array}
\right), \quad
\vec{t} =  \vec{0}.
\]
We stress that we do not restrict the range of minimization 
in (\ref{eq:finiterate}) to
the depolarizing or the Pauli channels. 
The minimization is carried out over the set of all the qubit channels
in (\ref{eq:region}).
The FindMinimum function
in Mathematica 9.0 was used for the minimization.

In Fig.\ \ref{fig1}, the key rates for various depolarizing rates
are plotted, and we compare key rates by our proposal and the
asymptotic rates by Matsumoto \cite{matsumoto13b92}.
We can observe that positive key rate is achieved at depolarizing
rate 6.4\% with $10^8$ samples.
The sample size refers to the total number $m$ of transmitted
qubits from Alice to Bob, including
qubits giving measurement outcomes $F_{\bar{0}}$ and $F_{\bar{0}}$
and qubits becoming sifted key.
In Fig.\ \ref{fig2}, the key rates for various sample sizes
are plotted with a fixed depolarizing rate 5\%,
and we also compare
key rates by our proposal and the
asymptotic rates by Matsumoto \cite{matsumoto13b92}.
We can observe that our key rates converge to the asymptotic one.
We only considered $\alpha=0.39$ and did not optimize
the value of $\alpha$ in Eq.\ (\ref{eq:alpha}).
The value $\alpha=0.39$ was also used in \cite{matsumoto13b92}.
$r_{pub}$ was always set to $0.5$ in our numerical computation.

\section{Conclusion}
Before this research, the secure key rate of the B92 quantum
key distribution protocol had not been reported.
We have clarified it.
Our analysis is based on the finite key rate formula
proposed by Scarani and Renner \cite{scarani08prl}
combined with our previous researches \cite{matsumoto13b92,sano10}.
We have shown that one can have a positive key
rate with $10^8$ samples over a depolarizing channel with
depolarizing rate 6.4\%.

\section*{Acknowledgment}
The authors would like to thank Prof.\ Shun Watanabe
for his helpful comments.
This research is partly supported by the NICT and the JSPS Grants
 No.\ 23246071.
This research is in part carried out during the second
author's stay in Aalborg university that was supported
by the                                                                   
 Villum Foundation through their VELUX Visiting Professor Programme             
 2013--2014.



\end{document}